\documentstyle[12pt,epsfig]{article}
\author{Juli\'an Candia$^{a}$ and Ezequiel V. Albano$^{b}$\\{}\\
$^a${\small\it Departamento de F\'{\i}sica, UNLP, 
CC67, 1900 La Plata, Argentina}\\
$^b${\small\it Instituto de Investigaciones Fisicoqu\'{\i}micas
Te\'{o}ricas y Aplicadas}\\{\small\it (INIFTA), UNLP, CONICET, 
Suc.4, CC16, 1900 La Plata, Argentina}}

\title{Non-equilibrium wetting transition in a
magnetic Eden model}

\begin{document}
\maketitle

\begin{abstract}
Magnetic Eden clusters with ferromagnetic interaction
between nearest-neighbor spins are grown in a confined
2d-geometry with short range magnetic fields acting on
the surfaces. The change of the growing interface curvature driven
by the field and the temperature is identified as a non-equilibrium wetting
transition and the corresponding phase diagram is evaluated.
\end{abstract}

\section{Introduction}

The study of irreversible growth models is a subject that
has attracted growing attention during the last decades.
Nowadays, this interdisciplinary field has experienced
a rapid progress due to both, their interest in many subfields
of physics, chemistry and biology, as well as by their relevance
in numerous technological applications. Recent progress in our
understanding of growth phenomena, with special emphasis on
the properties of rough interfaces, has been extensively
reviewed \cite{fam,bar,shl1,shl2,mar}.
On the other hand, the interaction of a bulk phase of a
system with a wall or a substrate may result in the
occurrence of very interesting wetting phenomena.
Wetting transitions have been experimentally observed and
theoretically studied in a great variety of systems in
thermal equilibrium, for reviews see e.g.\cite{die,par}.
In contrast, the study of wetting phenomena under non-equilibrium
conditions has, so far, received much less attention.
Within this context, very recently Hinrichsen et al. \cite{hin} have
introduced a non-equilibrium growth model of a one-dimensional
interface interacting with a substrate. The interface evolves
via adsorption-desorption processes which depart from detailed
balance. Changing the relative rates of these processes, a transition
from a binding to a non-binding phase is reported \cite{hin}. In fact, in
the study of wetting phenomena under equilibrium conditions,
wetting transitions are usually associated to the onset of unbinding
of an interface from a wall \cite{eva}.
The aim of this work is to study the properties of a
non-equilibrium wetting
transition which takes place in a variant of the irreversible
Eden growth model\cite{ede}, where the particles are replaced by spins
which may adopt two different orientations.
Such model is known as the
Magnetic Eden Model (MEM) and has been proposed by
Ausloos et al. \cite{mem1}. The MEM has originally been motivated 
by the study of the structural properties of magnetically 
textured materials \cite{mem1}. In the present work, the growing 
system  is
confined between two parallel walls where short range
boundary magnetic fields interact with the spins. Our investigation of
the properties of the MEM in such stripped geometry is also
motivated by recent experiments where the growth of 
quasi-one-dimensional strips of Fe on a Cu(111) vicinal surface  
has been studied \cite{iron}. Also, in a related context, the
study of the growth of metallic multilayers have shown a rich 
variety of new physical phenomena. Particularly, the growth
of magnetic layers of Ni and Co separated by a Cu spacer
layer has recently been studied \cite{cobre}. In this case,
the interaction between magnetic atoms in the bulk of the
respective layer may be different than that of such atoms
at the surface in contact with the Cu layer. Such interaction
may, in principle, be modeled by introducing a short
range boundary magnetic field, as we have proposed 
in the present work.     

\section{The model and the \\ Monte Carlo simulation method}

In the classical Eden model \cite{ede} on the square lattice,
the growth process starts by adding particles at the
immediate neighborhood (the
perimeter) of a seed particle. Subsequently, particles are
sticked at random to perimeter sites
leading to the formation of compact clusters with a self-affine
interface \cite{bar,shl1,shl2}.
The magnetic Eden model (MEM) \cite{mem1} considers an additional
degree of freedom due to the spin of the growing particles.
While early studies of the MEM have been performed using
a single seed placed at the center of the sample \cite{mem1},
for the purposes of the present work we have adopted a
different geometry. In fact, we have studied the MEM on
the square lattice in a rectangular (or stripped) geometry
of $L \times M$ ($1 \leq i \leq M$, $1 \leq j \leq L$) with $L \gg M$. 
The seed is a column
of $L$ spins located at $i = 1$ and cluster growth takes place
in one direction only, say for $i \geq 2$. Open boundary conditions
are also considered. A surface magnetic field $H$, acting on
the sites placed at $j = 1$ and $j = L$, accounts for the interaction
between the walls and the spins. It is assumed that each
spin $S_{ij}$ may adopt two possible orientations, namely up and
down (i.e., $S_{ij} = \pm 1$). Clusters are grown by selectively adding
spins at perimeter sites, which are defined as the
nearest-neighbor (NN) empty sites of the already occupied ones.
Considering a ferromagnetic interaction of strength $J > 0$ 
between NN spins, the
energy $E$ of a given configuration of spins is taken to be

\begin{equation}
E = - (J/2) \sum_{<ij,i^{'}j^{'}>} S_{ij}S_{i^{'}j^{'}} - 
H \sum_{<i,S>}(S_{i1}+S_{iL}) ,   
\end{equation}
  
\noindent where $<ij,i^{'}j^{'}>$ means 
that the summation in the first term is 
taken over occupied NN sites and $<i,S>$ denotes that the 
second summation has to be taken over occupied sites on 
both surfaces.
Thus, measuring the absolute temperature in
units of $J$ (the Boltzmann constant is set to unity), 
and the energy  and the surface 
magnetic field in units of $J$, the
change of energy $\Delta E$ involved in the addition of a spin 
$S_{ij}$ to the system is given by

\begin{equation}
\Delta E/T = -(1/T) S_{ij} \sum_{<ij,i^{'}j^{'}>} S_{i^{'}j^{'}} - 
(H/T)  (S_{ij} \delta_{j1} + S_{ij} \delta_{jL}) ,
\end{equation}

\noindent where the summation 
$<ij,i^{'}j^{'}>$ is taken over occupied NN sites keeping
$i,j$ fixed, and $\delta_{j1}$ and $\delta_{jL}$  are standard
Kronecker delta symbols. Therefore, 
the probability of a perimeter site to be occupied by a spin $S_{ij}$
is proportional to the Boltzmann factor exp$(-\Delta E /T)$, 
where $\Delta E$
is given by equation (2). At each step, all perimeter
sites are considered and the probabilities of adding up and down
spins have to be evaluated. After proper normalization of the
probabilities the growing site and the orientation of the spin
are determined through a pseudo-random number generator.
It is worth mentioning that while both the Hamiltonian and the
Boltzmann probability distribution considered for the MEM are
the same as the ones used for the Ising model in a rectangular
geometry with surface magnetic fields \cite{eva,eva1}, there exists an
essential difference between both models: namely, while the Ising
model deals with reversible spin configurations in thermodynamic
equilibrium, the MEM corresponds to a far-from-equilibrium
irreversible growth model.
Therefore, once the bulk of the aggregate is filled it becomes frozen
(i.e., it can not be modified any more due to further addition of spins).
This property allowed us to use a very well known
efficient simulation algorithm
which periodically removes the frozen part of the aggregate and only
keeps track of the active growing interface. In this way one
saves computer memory and large aggregates can be studied. In the
present work we have used strips of widths $L = 32$ and $L = 64$, and lengths
as large as $M = 3 \times 10^7$, generating spin aggregates 
of up to $10^9$ particles.
In order to quantitatively characterize the behavior of the system
we have measured the average magnetization of the frozen columns,
given by

\begin{equation}
m(i,L,T,H) = (1/L) \sum_{j = 1}^{L} S_{ij} ,                                      
\end{equation}

\noindent which plays the role of an order parameter. Also, the probability
distribution of the order parameter 
$P_{L}(m,T,H)$ has been evaluated \cite{kurt}.

\section{Results and discussion}

It is worth mentioning that we
have restricted ourselves to the $H \geq 0$ case without
loosing generality. In fact, first we have checked that the 
magnetic Eden growth process in a confined geometry is characterized
by an initial transient followed by a non-equilibrium stationary
state that is independent from the starting seed. So,
we have particularly employed an initial seed entirely
constituted by up spins throughout.
Thus, changing the sign of the applied field 
($H \rightarrow -H$) corresponds
to invert spin orientation at every lattice site. Then,
the order parameter probability distribution can be simply obtained
by replacing $P_{L}(m) \rightarrow P_{L}(-m)$. 
Analogous replacements also hold
for other observables that can be computed from $P_{L}(m)$, such
as the average of the absolute column magnetization defined
below (see equation 4).

\begin{figure}
\centerline{{\epsfysize=2.3in \epsffile{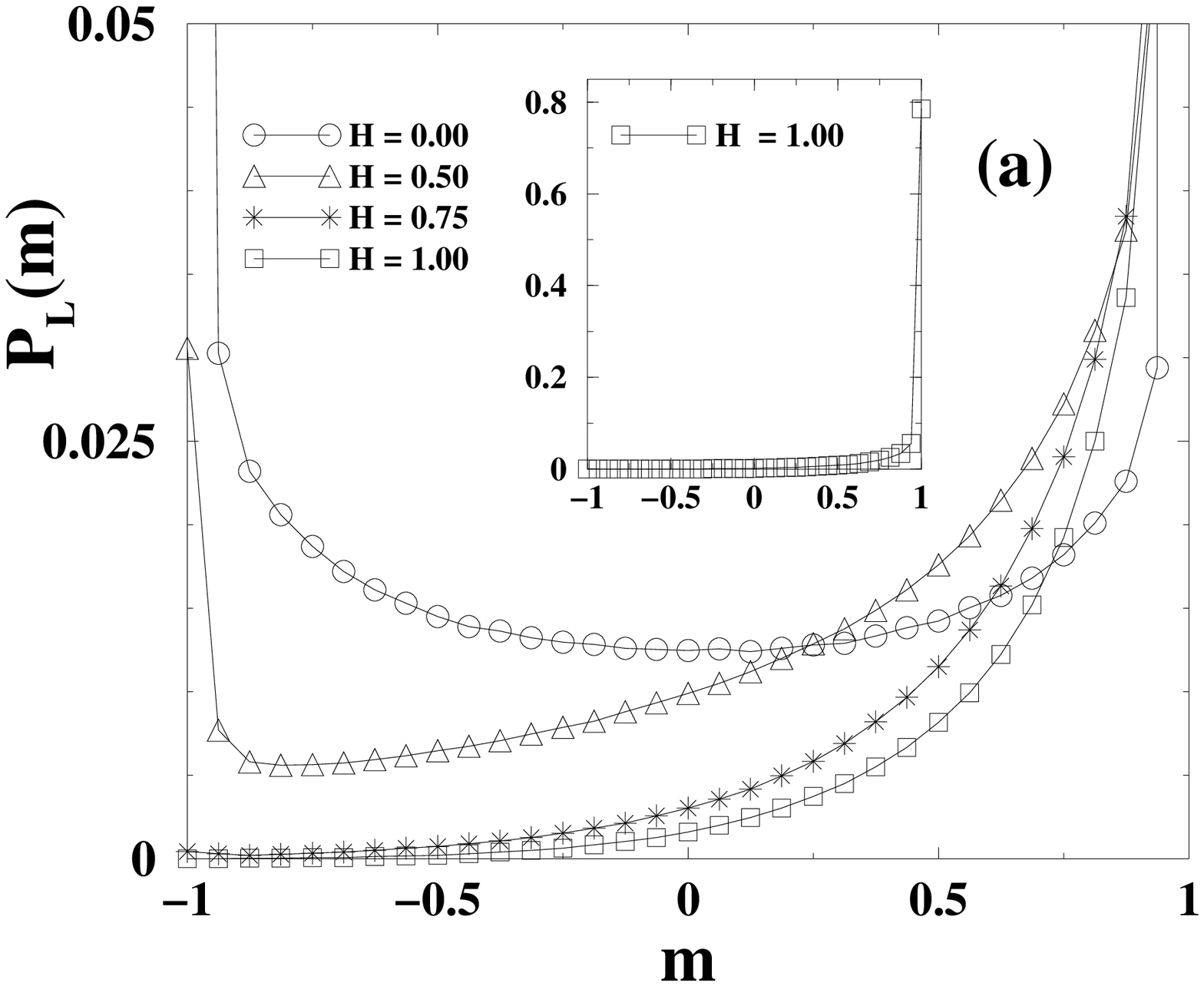}}}
\centerline{{\epsfysize=2.3in \epsffile{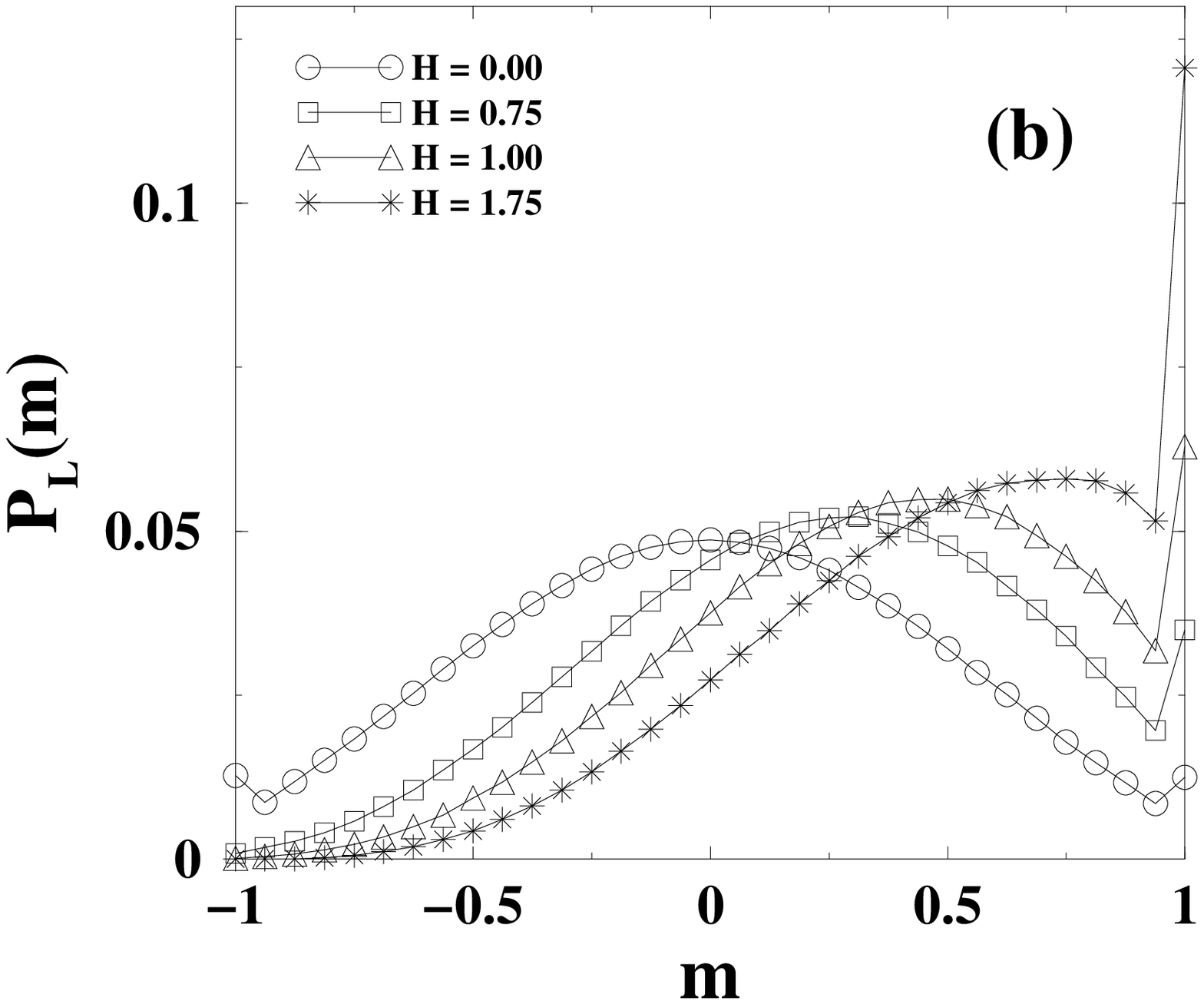}}}
\centerline{{\epsfysize=2.3in \epsffile{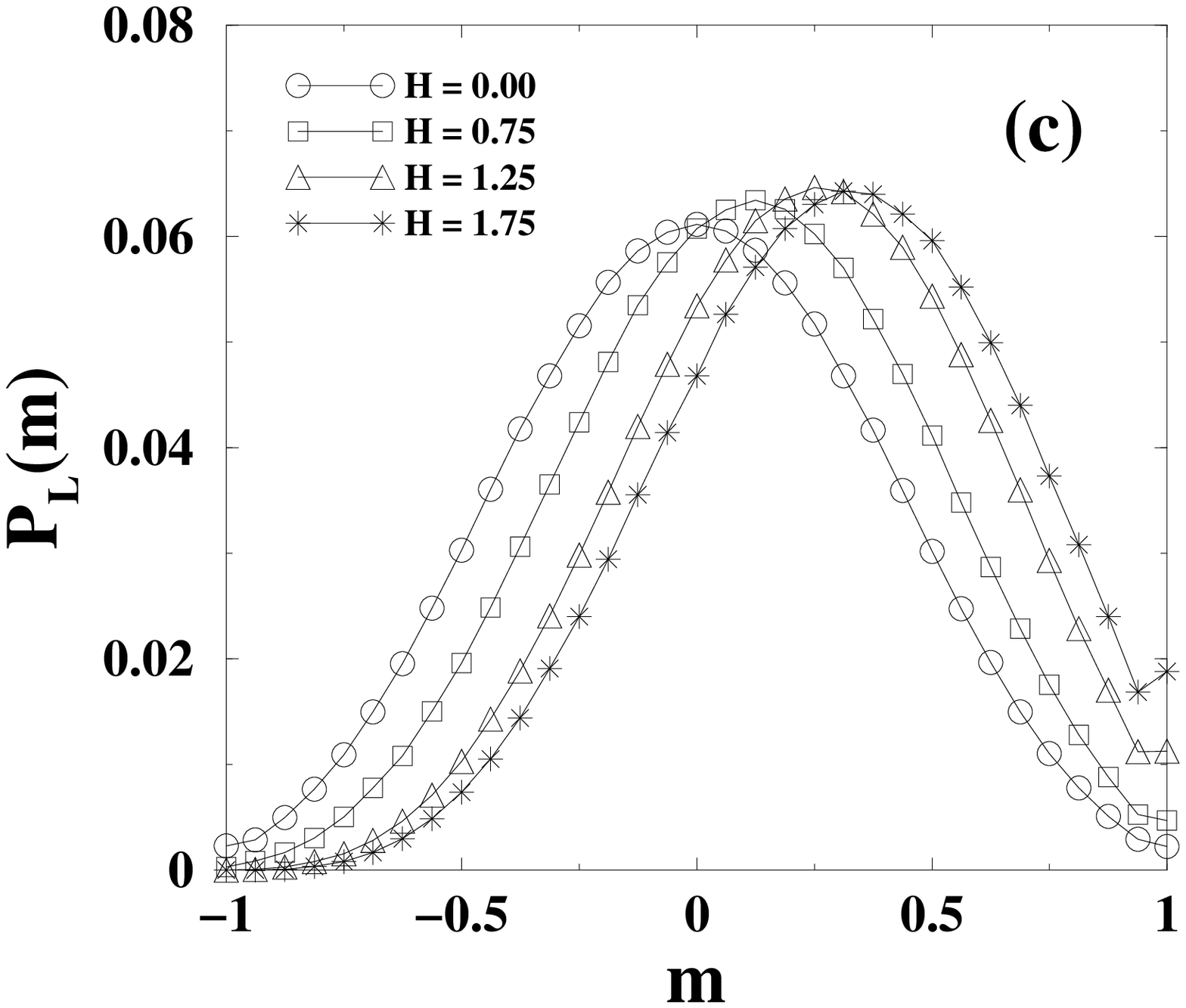}}}
\caption{Plots of the order parameter distribution function
$P_{L}(m)$ vs m obtained for different values of $H$ as 
indicated in the figures. (a) Results for $T = 0.5$. The vertical 
axis has been truncated in order to allow a detailed observation 
of the dependence of 
$P_{L}(m)$ with m. The inset shows a plot of $P_{L}(m)$ vs $m$ 
obtained taking $H = 1.0$, where the sharp peak at $m = 1$ 
can be observed. (b) and (c) show results obtained for $T = 0.8$ and 
$T = 1.0$, respectively. More details in the text.}
\label{FIG. 1}
\end{figure}

Figure 1 shows typical plots of the probability distribution
of the order parameter, as obtained for different temperatures
and fields (unless otherwise stated, we consider the case
of strip width  $L = 32$ throughout). 
Since the surface field is always assumed to be
positive, all distributions are biased towards positive values of m.
Figure 1(a) shows a typical low-temperature distribution that
corresponds to $T = 0.5$. There one observes that
for weak fields two peaks of $P_{L}(m)$ clearly emerge at $m = \pm 1$.
Particularly, for $H = 0$ the distribution is symmetric and the 
average magnetization vanishes.
As naturally expected, for $H > 0$ the peak at $m = 1$ is higher due
to the applied surface fields. This result points out that
the system undergoes fluctuations, since spin columns happen to be
mainly builded up by parallel-aligned spins with a
single orientation, either up or down.
Increasing the field ($H \geq 0.75$) the negative 
peak of $P_{L}(m)$ vanishes
showing that the surface field is strong enough in order to
suppress such fluctuations.
At $T = 0.8$ (figure 1(b)), $P_{L}(m)$ is strongly biased by the field
and only small peaks at $m = -1$ can be observed for very weak fields
($H < 0.25$). It should also be noted that the curvature of $P_{L}(m)$
changes, as compared to figure 1(a). In fact, at $T = 0.8$ and for
weak fields the amount of disorder in the aggregate is large
enough so that  $P_{L}(m)$ becomes peaked around $m \simeq 0$
and the average magnetization is close to zero.
The Gaussian-like shape of the distribution curves for $H < 0.5$
becomes distorted by the effect of the field causing a shift
of the whole distribution towards larger values of m as well
as the occurrence of a sharper peak at $m = 1$, that grows with
increasing the applied field and becomes dominant for $H \geq 1$.
For higher temperatures ($T = 1.0$ in figure 1(c)) the Gaussian shape
for weak fields can clearly be observed, while the bias caused
by the field has minor influence as compared with the 
former cases (figures 1(a-b)).
Due to the observed fluctuations of $m(T,L,H)$, the order parameter
as defined by equation (3) will tend to vanish upon averaging
over all frozen columns. Therefore, in order to avoid this
effect, it is convenient to redefine the order parameter as
the average of the absolute column magnetization \cite{eva}, i.e.

\begin{equation}
<|m(L,T,H)|> = (1/M^{*}) \sum_{i = 1}^{M^{*}} |m(i,L,T,H)|   ,
\end{equation}

\noindent where $M^{*} < M$ is the number 
of frozen columns where the growing
process has definitively stopped, that is, the number of completely
filled columns.

\begin{figure}
\centerline{{\epsfysize=2.3in \epsffile{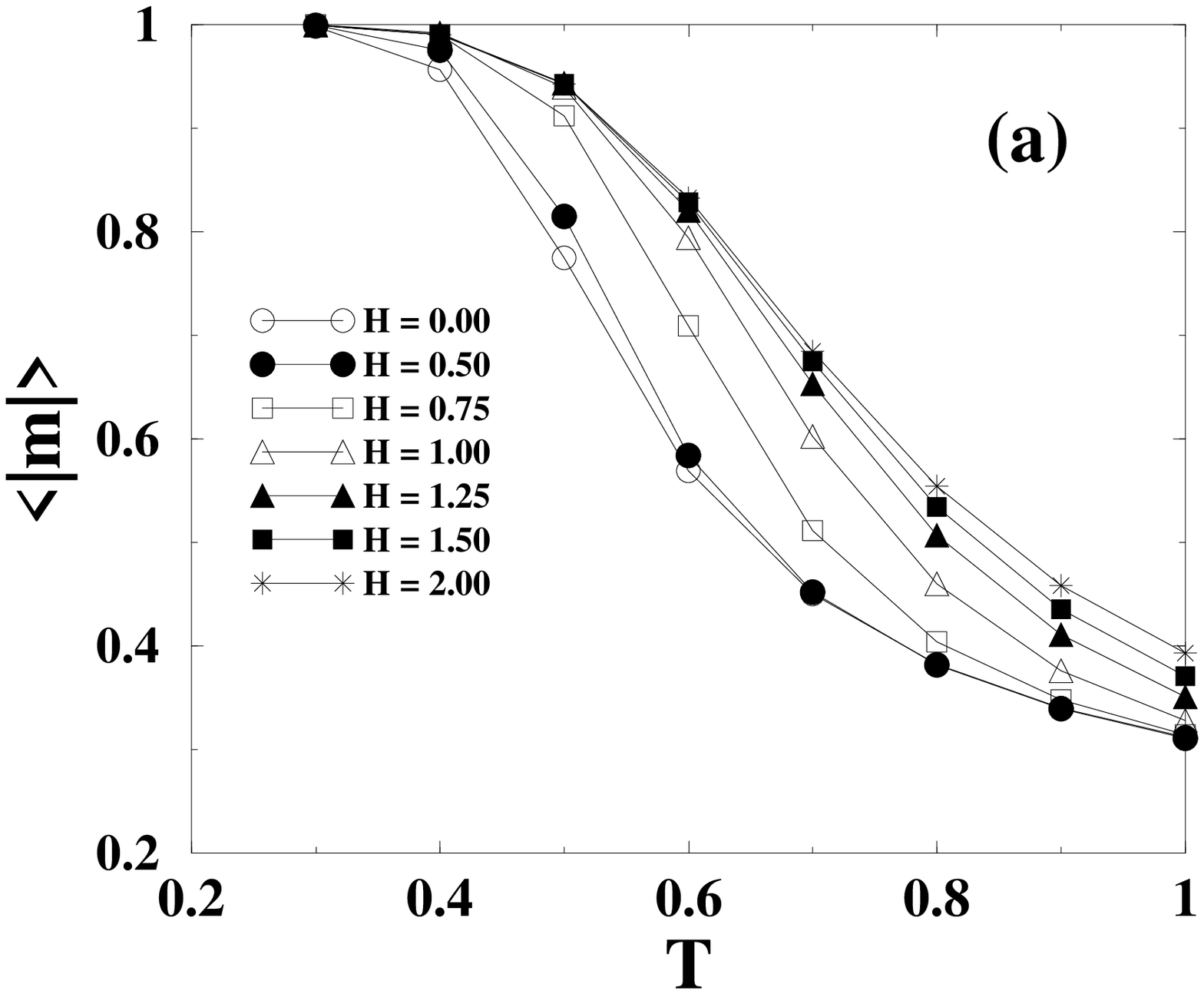}}}
\centerline{{\epsfysize=2.3in \epsffile{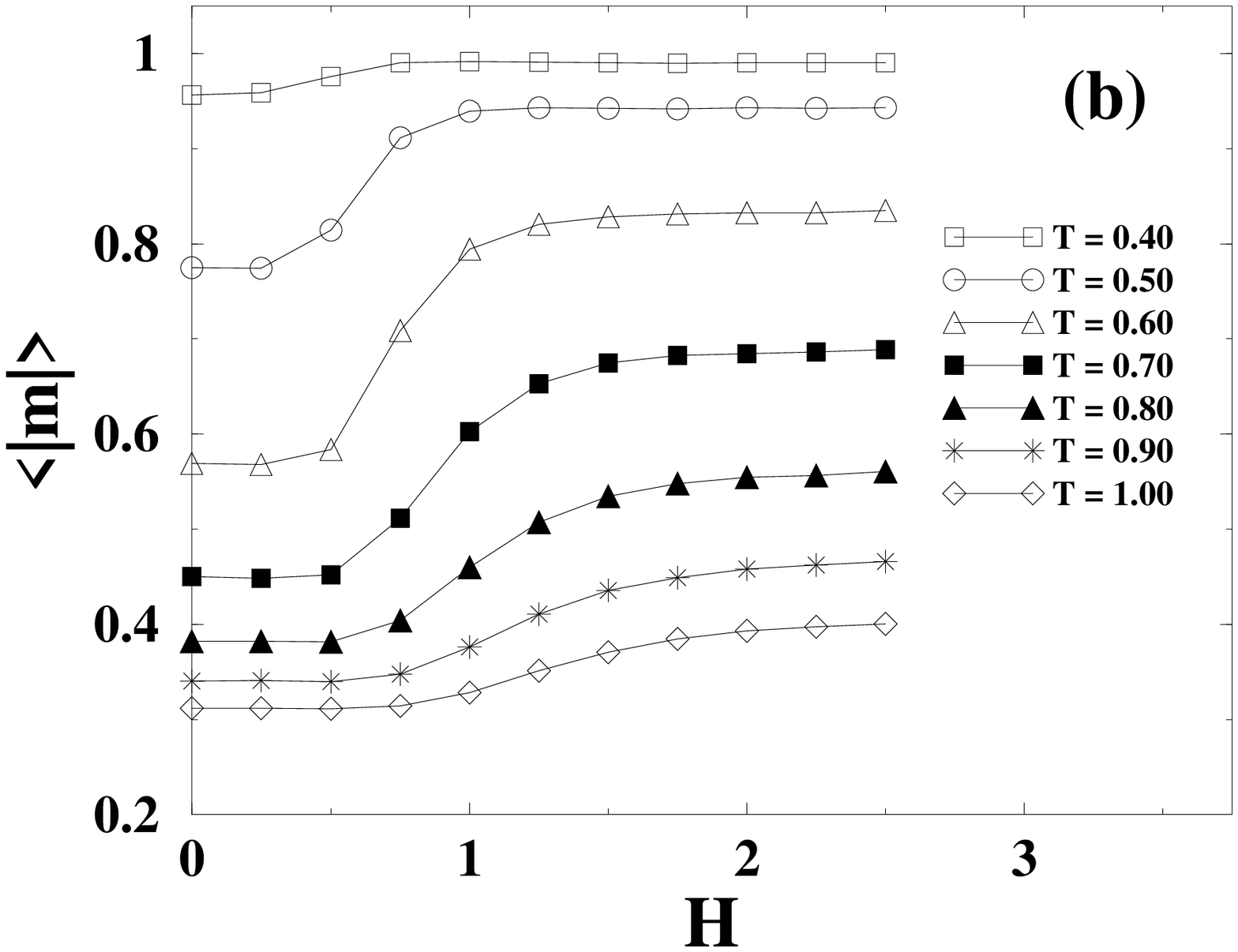}}}
\caption{(a) Plots of the order parameter versus
the temperature 
obtained for different values of the surface magnetic 
field as indicated in the figure.
(b) Isotherms showing the dependence of the order parameter 
with the surface magnetic field, obtained for the 
temperatures indicated in the figure.}
\label{FIG. 2}
\end{figure}

Figure 2(a) shows the dependence of $<|m(L,T,H)|>$ on the
temperature for different values of the field, while figure 2(b)
shows the plots of $<|m(L,T,H)|>$ vs. $H$ obtained at different
fixed temperatures. At low temperatures (say $T < 0.4$) and even for
very weak surface fields, the growth of magnetic Eden aggregates
with chiefly parallel-oriented spins is observed. The absolute
magnetization (and consequently the order) also remains quite
large even when temperature is increased up to $T = 1$ 
(figure 2(a)). When
comparing these plots with standard magnetic systems in
equilibrium, e.g. the Ising model with surface fields \cite{eva}, it
is clear that for the lattices used in this work the MEM order-disorder
transition is strongly rounded due to finite-size effects.
The isotherms of figure 2(b) show that for each studied
temperature there exists a surface magnetic field capable of
causing the saturation of $<|m|>$.
Of course, the dependence of the order parameter on the surface
field at fixed temperature is smooth, in contrast with the jumps
observed in the Ising system. This is not only attributable to
finite-size effects. Actually, the fact that the magnetic fields
are only applied to the boundaries of the aggregate (but not to
the whole bulk of spins) plays a major role.

\begin{figure}
\centerline{{\epsfysize=2.7in \epsffile{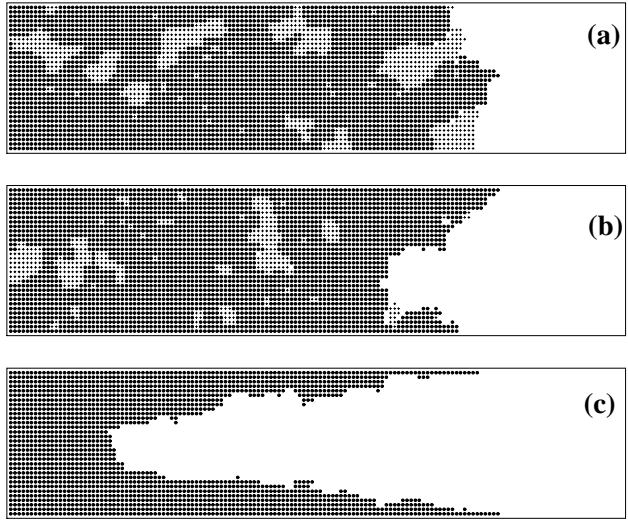}}}
\caption{Snapshot showing typical configurations of 
magnetic Eden aggregates. Filled circles and crosses correspond to 
spins pointing up (parallel to the magnetic 
surface field) and spins pointing down, respectively.
The lattice width is $L = 32$. (a) $T = 0.67$ and $H = 0.0$.
(b) $T = 0.67$  and $H = 1.33$. (c) $T = 0.33$ and $H = 1.33$.}
\label{FIG. 3}
\end{figure}

\indent Figure 3 shows typical MEM snapshot configurations obtained at
different temperatures and surface fields.
The different shapes of the growing interfaces observed in figure
3 can be understood on the base of simple arguments. For $H = 0$ and
due to the fact that open boundary conditions are imposed at $j = 1$
and $j = L$, empty perimeter sites at the walls of the sample will
experience a mixing neighbor effect, that is, the average number
of NN occupied sites will be lower than for the case of perimeter sites
on the bulk. Consequently, the system will preferentially grow
along the center of the sample as compared to the walls, and the
resulting growth interface will exhibit a convex shape (figure 3(a)).
For $H > 0$, the growing probability of perimeter sites at the
walls of the system will be favored by an additional probabilistic
factor given by exp$(\pm  H/T)$, as it follows from equation (2).
If $H/T$ becomes large enough, the preferential growth along
the walls will dominate (figures 3(b) and 3(c)) and the interface curvature
will become concave. So, from a qualitative point of view, the
figures 3(a,b,c) allow us to expect the occurrence of a convex-concave
transition in the curvature of the growth interface. Such transition
is identified as a non-equilibrium magnetic wetting transition, since
a concave interface wets the walls while they remain dry when
the interface grows with convex curvature.

\begin{figure}
\centerline{{\epsfysize=2.4in \epsffile{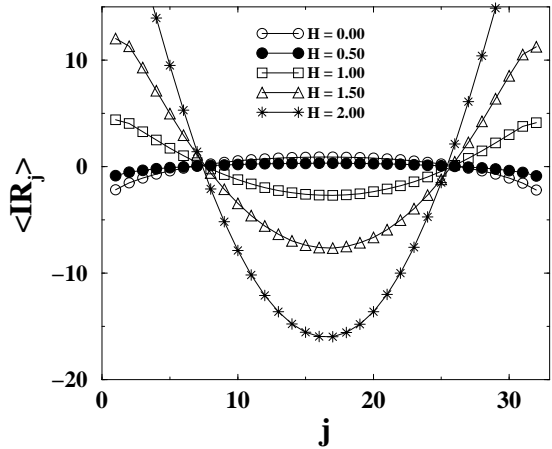}}}
\caption{Plots of the averaged interface profile versus $j$ obtained 
for $T = 0.7$ and different values of the surface magnetic field 
as indicated in the figure. The plot corresponding to $H = 2.0$
has been truncated in order to allow a detailed observation of the 
profile's curvature for lower $H$ values.}
\label{FIG. 4}
\end{figure}

\indent In order to perform a quantitative study of the wetting transition
it is convenient to define the location and the curvature of the
growing interface. We assume that each row of the system
contributes with the outermost perimeter site (i.e., the one with
the largest value of the longitudinal $i-$th coordinate, for a given row
number $j$) to the growing interface. Let $I_{j}(t)$ be the $i-$th abscissa
corresponding to the $j-$th row at time $t$. Then, the interface 
center of mass, that we take as the location of the interface 
at time $t$, $I(t)$, is given by

\begin{equation}
I(t) = (1/L) \sum_{j = 1}^{L} I_{j}(t) .                                  
\end{equation}

\noindent Subsequently one can evaluate the coordinates of the interface
relative to its center of mass location at time $t$, namely
$IR_{j}(t) = I_{j}(t) - I(t)$, $j = 1, 2, ..., L$. 
In this way we obtain a set
\{ $IR_{j}(t)$ \} that appropriately describes the interface at any time $t$
during the growing process. In order to increase the statistics,
we may evaluate the average relative interface $<IR_{j}>$ given by

\begin{equation}
<IR_{j}> = \{1 / (t_{f}-t_{i}+1)\} \sum_{t = t_{i}}^{t_{f}} IR_{j}(t)                      
\end{equation}

\noindent where we take into account interface coordinates measured at
different times between $t_{i}$ and $t_{f}$.
\begin{figure}
\centerline{{\epsfysize=2.4in \epsffile{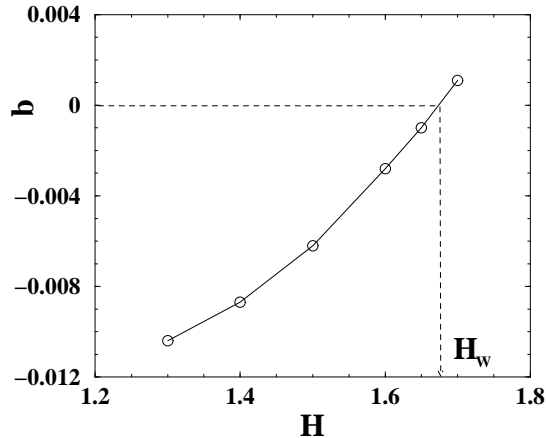}}}
\caption{Plot of $b$ versus $H$ obtained for $T = 2.00$. The dotted line
shows the point where $b$ changes its sign allowing us to 
identify the critical wetting field ($H_{w}$), 
as indicated in the figure. More details in the text.}
\label{FIG. 5}
\end{figure}

\begin{figure}
\centerline{{\epsfysize=2.5in \epsffile{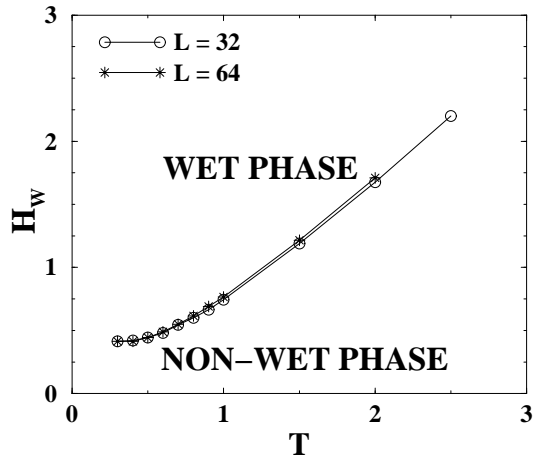}}}
\caption{Wetting phase diagram, $H_{w}$ vs $T$, showing the critical
line at the boundary between the wet and non-wet phases. 
Results obtained for $L = 32$ and $L = 64$, as indicated in the figure.}
\label{FIG. 6}
\end{figure}

Figure 4 shows plots of $<IR_{j}>$ vs. $j$ corresponding to $T = 0.7$ for
different values of the surface field $H$. There, it becomes evident
how the applied field drives the convex-concave curvature change.
The curved interfaces have been fitted by means of a fourth-degree
polynomial given by $p(j) = a + b(j-j_c)^{2}+ c(j-j_c)^{4}$, 
where $j_c = (L+1)/2$. 
All fits were
characterized by a dominant quadratic term which defined the
interface curvature and a practically negligible quartic coefficient.
Thus, the sign change of the quadratic coefficient $b$ allows the
identification of the convex-concave curvature transition: for
$b > 0$ the interface is concave and the cluster wets the walls, while
for $b < 0$ it is convex and the walls remain dry. So, given a fixed
temperature $T$, the magnetic field at the wetting transition $H_{w}$ is
the one that corresponds to $b = 0$. Figure 5 shows a plot of $b$ vs. $H$
obtained for $T = 2.0$, where the change of sign of $b$ 
can clearly be observed.
In this example, by means of a linear interpolation, the value 
$H_{w} = 1.67 \pm  0.03$ is obtained.

\indent Following this procedure, we can quantitatively obtain the wetting
phase diagram $H_{w}$ vs. $T$. These results are shown in figure 6, which
corresponds to strip widths $L = 32$ and $L = 64$. They also suggest that
the location of the critical wet non-wet curve is only weakly sensitive
to finite-size effects.
The monotonic growth of the $H_{w}$ vs. $T$ curves shown, reflects the fact
that a larger surface magnetic field is needed in order to stabilize
the thermal noise caused by higher temperatures.

\section{Conclusions}

\indent In this work, rectangular strips on the square lattice are used
to grow magnetic Eden clusters with ferromagnetic interactions
between nearest neighbor spins and short range magnetic fields
applied at the surfaces. For weak surface fields, the mixing
neighbor effect at the surface causes the growth of convex interfaces.
However, when the field is increased, the preferential growth
of spins along the surface turns the interface curvature concave.
Such curvature change has been rationalized in terms of a wetting transition,
and the corresponding wet non-wet phase diagram has been evaluated.
We expect that the present study will stimulate further work
in the field of non-equilibrium wetting transitions, a topic
of widespread technological and scientific interest which has
remained almost unexplored till the present.

\section*{Acknowledgments}

This work is supported  financially  by CONICET, UNLP, CIC (Bs. As.), 
ANPCyT and Fundaci\'on Antorchas (Argentina) and the Volkswagen 
Foundation (Germany).

\end{document}